\documentclass{ws-ijmpb}

\usepackage{amsmath}
\usepackage{amstext}
\usepackage{amssymb}
\usepackage{color}
\usepackage[super]{cite}

\begin{document}
\newcommand{\onlinecite}[1]{\hspace{-1 ex} \nocite{#1}\citenum{#1}} 

\markboth{M. Przedborski, S. Sen, T. Harroun}
{Long-term behaviour of Hertzian chains between fixed walls is really equilibrium}

%
\catchline{}{}{}{}{}
%

\title{Long-term behaviour of Hertzian chains between fixed walls is really equilibrium}

\author{Michelle Przedborski\footnote{mp06lj@brocku.ca}}
\address{Department of Physics, Brock University, 1812 Sir Isaac Brock Way, \\ St. Catharines, Ontario, Canada L2S 3A1}

\author{Surajit Sen\footnote{sen@buffalo.edu}}
\address{Department of Physics, State University of New York, Buffalo, New York \\ 14260-1500, USA}

\author{Thad A. Harroun\footnote{thad.harroun@brocku.ca}}
\address{Department of Physics, Brock University, 1812 Sir Isaac Brock Way, \\ St. Catharines, Ontario, Canada L2S 3A1}

\date{\today}
\begin{history}
\received{\today}
\revised{Day Month Year}
\end{history}

\begin{abstract}
We examine the long-term behaviour of non-integrable, energy-conserved, 1D systems of macroscopic grains interacting via a contact-only generalized Hertz potential and held between stationary walls. Existing dynamical studies showed the absence of energy equipartitioning in such systems, hence their long-term dynamics was described as \emph{quasi-equilibrium}.
Here we show that these systems do in fact reach thermal equilibrium at sufficiently long times, as indicated by the calculated heat capacity. This phase is described by equilibrium statistical mechanics, opening up the possibility that the machinery of non-equilibrium statistical mechanics may be used to understand the behaviour of these systems away from equilibrium. \end{abstract}

\keywords{Hertz chain; solitary wave; equilibrium}
\maketitle

\section{Introduction}
In recent years, 1D systems of discrete macroscopic grains interacting via a power-law contact potential and held between fixed walls have attracted considerable attention,\cite{Nesterenko1983,Nesterenko1985,Nesterenko1995,Sinkovits1995,Sen1996,Coste1997,Sen1998,Chatterjee1999,Hinch1999,Hong1999,Ji1999,Manciu1999,Hascoet2000,Sen2001,Nesterenko2001,Rosas2003,Rosas2004,Nesterenko2005,Job2007,Sokolow2007,Zhen2007,Herbold2009,Santibanez2011,Takato2012,Vitelli2012,Sen2001b,Nakagawa2003,Sokolow2005,Hong2005,Doney2006,Melo2006,Doney2009,Daraio2006,Job2009,Theocharis2009,Boechler2010,Theocharis2010,Breindel2011,Przedborski2015b,QEQ,Sen2005,Job2005,Avalos2007,Avalos2011,Avalos2014,SSWs,Sen2008,Przedborski2015} primarily because of their usefulness for a variety of applications related to shock mitigation\cite{Sen2001b,Nakagawa2003,Sokolow2005,Hong2005,Doney2006,Melo2006,Doney2009} and energy localization.\cite{Daraio2006,Job2009,Theocharis2009,Boechler2010,Theocharis2010,Breindel2011,Przedborski2015b} This is facilitated by their ability to support solitary wave (SW) propagation, which can be initiated in uncompressed chains by a simple edge impulse. However, unlike solitons found in continuum systems (which experience only a slight phase shift upon collision with another soliton), SWs in these discrete systems suffer from weak interactions with each other and with system boundaries. 

In particular, SWs are not perfectly preserved in these discrete, non-integrable systems since grains are capable of breaking contact, disrupting the SW's flow. Boundary collisions\cite{QEQ,Sen2005,Job2005}  result in the partial decimation of the original SW and the creation of much smaller magnitude secondary solitary waves\cite{Job2005,SSWs} (SSWs). In contrast, collisions of SW species with each other\cite{Avalos2007,Avalos2011} leads to energy being exchanged between waves, and thus a potential for the increase in energy amplitude of one of the waves. Many collisions between SW species in a system with zero energy dissipation therefore leads to both breakdown and buildup processes of SSWs. 

For singular perturbations, these breakdown and buildup processes lead the system after a long time to an equilibrium-like, ergodic phase.\cite{QEQ,Sen2005,Avalos2007,Avalos2011,Avalos2014,Sen2008} This spatially-symmetric phase is attained when the rates of SSW formation and breakdown balance, and is marked by a large number of SSWs that are equally likely to be moving in either direction. For sufficiently strong and unique perturbations, unusually large~\cite{QEQ,Sen2005,Avalos2007,Avalos2011,Avalos2014,Sen2008} and occasionally persistent (rogue)\cite{Han2014} fluctuations in the system's kinetic energy are seen at long times. This impedes an equal sharing of energy among all grains in the system, hence the long-term dynamics of 1D systems of interacting grains has been described as quasi-equilibrium (QEQ).\cite{QEQ,Sen2005,Avalos2007,Avalos2011,Avalos2014,Sen2008} 

To the time scales previously considered in dynamical studies, QEQ was observed to be a general feature of systems with no sound propagation.\cite{QEQ,Avalos2011}
However, recent work has addressed whether QEQ is the final state for such systems.\cite{Onorato2015,Przedborski2016} These studies found that thermalization is indeed possible after very long times, and that the time scale to equilibrium increases with the degree of nonlinearity. While the relaxation to equilibrium was inaccessible when the first numerical experiments were performed,~\cite{Fermi1955} it can now be probed thanks to current technology. This solves a long outstanding problem regarding the long-time evolution of these strongly nonlinear, discrete, non-integrable systems. 

Here we show, for the first time, that the very long-time dynamics of Hertzian chains is described by the equilibrium statistical mechanics of a microcanonical ($\bm{NVE}$) ensemble of interacting particles. 
We accomplish this primarily by illustrating that at sufficiently long times, energy is indeed equipartitiond among the \textit{independent degrees of freedom}, as indicated by the calculated finite heat capacity. 

The remainder of the paper is organized as follows. In Sec.~\ref{sec:Model} we introduce the model for the Hertzian chains, and derive the associated prediction for the equilibrium value of the heat capacity. Then we give the details of the simulation parameters, and in Sec.~\ref{sec:calculation} the details of numerical calculations. In Sec.~\ref{sec:results} we compare MD data with the predicted equilibrium values to establish that our systems do equilibrate at long times. Finally, we give some concluding remarks and discuss future research directions in Sec.~\ref{sec:conclusions}. 

\section{\label{sec:Model}Model and Simulations}
The specific systems under consideration are 1D chains of $N$ grains, each with mass $m$ and radius $R$, interacting via a Hertz-like contact-only potential.~\cite{Hertz1882} The Hamiltonian describing the system is the sum of a kinetic energy term $K$ and potential energy term $U$ associated with grain interactions, given by
\begin{equation}
\mathbf{H} = K + U = \frac{1}{2} \sum_{i=1}^N m v_i^2 + \sum_{i=1}^{N-1} a \Delta_{i,i+1}^n,
\label{eq:hamiltonian}
\end{equation}
where $\bm{v}_i$ is the velocity of grain $i$ and $\Delta_{i,i+1} \equiv 2R - (x_{i+1}-x_i) \ge 0$ is the overlap between neighbouring grains, located at position $x_i$. If $\Delta_{i,i+1}<0$, there is no potential interaction. In the above expression, the exponent $n$ is shape-dependant ($n=2.5$ for grains with ellipsoidal contact geometries, such as spheres), and $a$ contains the material properties of the grains.\cite{Sun2011} The grain interactions with the fixed walls adds two terms to the Hamiltonian.\cite{Przedborski2015} For homogeneous systems in which the grains and walls are comprised of the same material, such as the ones considered here, the coefficient describing the grain-wall interaction $a_w$ is related to the grain-grain interaction coefficient $a_g$ via $a_w = \sqrt{2} a_g$.\cite{Przedborski2015} Further details on how the material properties affect the transition from the non-ergodic SW phase to the QEQ phase can be found in Refs.~\onlinecite{Przedborski2015b} and~\onlinecite{Przedborski2015}.

\begin{figure}[ht]
\centering
\includegraphics[width=0.95 \textwidth]{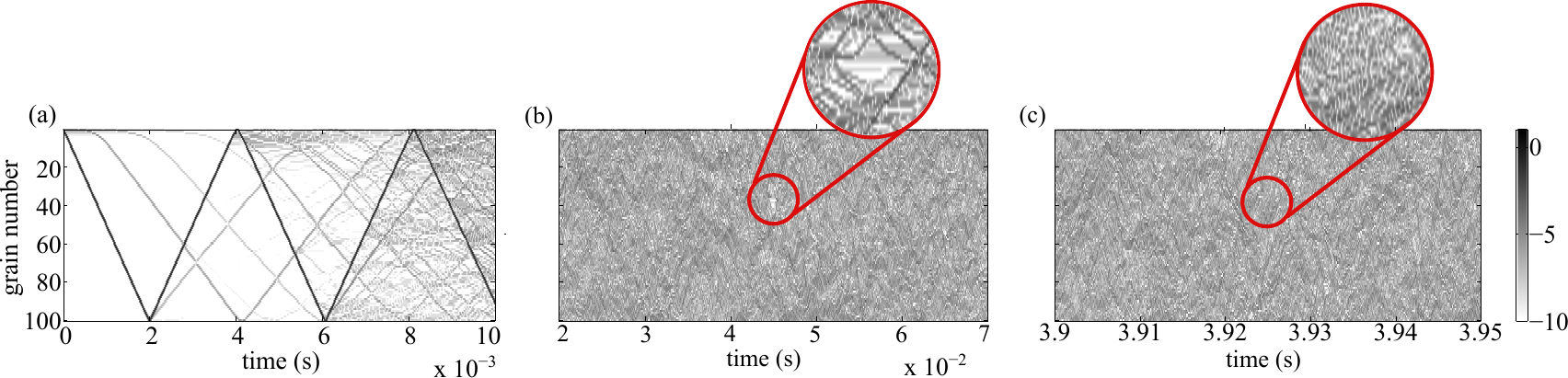}
\caption{(Color online) Kinetic energy density plots observed in systems described by Hamiltonian~(\ref{eq:hamiltonian}), and given a single initial edge impulse at $t=0$. This system corresponds to $N$=100, $n$=2.5. All data is normalized to the input energy and presented on a logarithmic scale. (a) The non-ergodic (SW) phase showing the breakdown of the initial SW and creation of SSWs; (b) The QEQ phase, illustrating large regions of hot and cold spots, hence large kinetic energy fluctuations; (c) The equilibrium phase. Comparing the inserts in (b) and (c), it is clear that the fluctuations have relaxed to smaller values in equilibrium. }
\label{fig:fig1}
\end{figure}

If at time $t=0$ such a system is given an edge impulse, a SW will propagate through the chain and eventually break down into a sea of secondary solitary waves (SSWs), as illustrated in Fig.~\ref{fig:fig1}. This process happens sufficiently long after the initial perturbation to the system and can be modelled as a transition from a non-ergodic (SW) phase to an ergodic (equilibrium) phase. Energy is, on average, shared equally among all the grains in this late-time phase since there will be a large number of SSWs traversing the system in either direction, each spanning several grains. For systems with energy dissipation turned off, a $\bm{NVE}$ ensemble is hence established. This means that the long-term dynamics of  Hertz chains is best described by the statistics of a 1D gas of interacting particles in thermodynamic equilibrium.

The QEQ phase bridges the slow transition from the SW phase to the equilibrium phase, in analogy to a coexistence region in a continuous phase transition. QEQ is distinct from equilibrium primarily because equipartitioning of energy among the degrees of freedom does not hold in the latter, as indicated by the unusually large fluctuations in the system kinetic energy, depicted in Fig.~\ref{fig:fig1}. As the system approaches equilibrium, the kinetic energy fluctuations relax to much smaller, but \emph{finite} values in finite systems. Thus the equal sharing of energy in finite systems happens only in an average sense, and each grain will not have \emph{exactly} the same kinetic energy at any instant in time. (Rather, the kinetic energy of each grain fluctuates according to the same probability density function.) 

Equal energy sharing is thus reflected in the value of the kinetic energy fluctuations. Moreover, since the system kinetic energy fluctuations are directly connected to the heat capacity in a $\bm{NVE}$ ensemble,\cite{Lebowitz1967} the latter provides an excellent way to probe the extent to which energy equipartitioning holds. To show that Hertzian systems ultimately move to an equilibrium phase where energy is being equipartitioned, we thus demonstrate agreement between calculated heat capacities from MD simulations and values predicted by Tolman's generalized equipartition theorem.\cite{Tolman1918} 
 
Since the generalized equipartition theorem applies to a canonical ensemble, the equilibrium value predicted for the specific heat of a Hertz chain is valid only in the thermodynamic limit, where we can rely on the equivalence of statistical ensembles. In this limit, we expect the specific heat per grain to be a constant. In a subsequent manuscript, we derive a correction term for finite system sizes.\cite{Przedborski2016}

In ergodic systems, Tolman's generalized equipartition theorem\cite{Tolman1918} applied to the Hamiltonian above results in an average total energy per grain $\left \langle \epsilon \right \rangle = k_B T/2 + k_B T/n$, where $k_B$ is Boltzmann's constant and $T$ is the \emph{canonical} temperature (and angular brackets denote an ensemble average, or equivalently, a time average). Taking a simple temperature derivative, the corresponding specific heat per grain is then
\begin{equation}
C_V^{\mathrm{Eq}} = \left ( \frac{n+2}{2n}\right ) k_B,
\label{eq:sp}
\end{equation}
which evidently depends \textit{only} upon the exponent in the potential, i.e. there is no dependence on grain (or wall) material, grain size, or temperature. The equivalence of different statistical ensembles when $N\to\infty$ then implies that Eq.~(\ref{eq:sp}) is the value of the specific heat in a $\bm{NVE}$ ensemble in this limit, which is expected when energy is equipartitioned. We compare Eq.~\ref{eq:sp} with the calculated specific heats obtained from numerical simulations to deduce the nature of the long-term dynamics of systems described by Hamiltonian~(\ref{eq:hamiltonian}). 

\subsection{\label{sec:simulations}Simulation details}
To examine the very long-time dynamics of Hertzian chains, we ran MD simulations of a 1D monatomic chain of $N$ grains held between fixed walls and described by the Hamiltonian in Eq.~(\ref{eq:hamiltonian}). To implement the fixed walls, we add two terms to the Hamiltonian,\cite{Przedborski2015} where the walls are taken to be grains of radius $R \to \infty$ to ensure that they do not move, while simultaneously relaxing the condition that they must be flat. Our grains and walls are made of steel, and the grains are 6~mm in radius. We do not apply any pre-compression, or squeezing of the chain in the set up, but rather each grain is initially just touching its neighbour between walls $N(2R)$ apart. 

We consider values of the potential exponent $n$ from 2 (harmonic) to 5, and system sizes from $N=10$ to 100. A standard velocity Verlet algorithm is used to integrate the equations of motion with a 10~ps timestep, and no dissipation is included. The grains are set into motion with either asymmetric edge perturbations (initial velocity given to the first grain only, directed into the chain, causing a single initial SW to propagate through the system); or with symmetric edge perturbations (initial velocity given to the first and last grain, both directed into the chain, and causing two initial SWs of equal magnitude to propagate toward the chain centre). In both cases, the initial SW(s) breaks down in collisions with boundaries (and with each other) and in the formation of gaps, creating numerous secondary solitary waves (SSWs). After a period of time, the number of SSWs increases to a point where the system enters into quasi-equilibrium.\cite{QEQ,Sen2005,Avalos2011,Avalos2014} We allow the system to evolve for a substantial amount of time past this phase change, and at least an order of magnitude longer than previous work has considered. The system energy is constant to nine decimal places for the entire simulation.

The length of time to reach equilibrium is primarily determined by the potential exponent $n$,\cite{Sen2008} so we adjust the velocity perturbation such that the system arrives at equilibrium in a reasonable computational time frame. To get an estimate for the optimal velocity perturbation, we utilized an $n=2.5$, initial velocity of $9.899\times10^{-5}$mm/$\mu$s simulation for reference. Equilibrium is reached in this system by the time $t=1$s. The relation between solitary wave speed $v_s(n)$ and impulse speed $v_i$ is given by\cite{Sen2008} $v_s(n) \sim v_i^{(n-2)/n}$. We use this with $v_i = 9.899\times10^{-5}$mm/$\mu$s to get a very rough estimate of how solitary wave speed varies with $n$ in these systems. The order of magnitude of the ratio $v_s(n)/v_s(n=5/2)$ gives an idea of the maximum factor by which the initial perturbation should be scaled for a higher value of $n$, if the system is to equilibrate in roughly the same amount of time.  

Of course, the velocity perturbation cannot be too large if the Hertz law is to remain valid,\cite{Hertz1882} so it was necessary in some cases to choose velocities that were quite a bit smaller than the maximum predicted by this estimate, and we therefore collected at least one second of real time data for $n=2, 2.5, 2.75$, and even longer (up to $6$~s) for larger values of $n$. These are much longer times than previously reported. Data of grain position and velocity are recorded to file every 1~$\mu$s, though we re-sample the data at time intervals beyond the dampening of velocity autocorrelation; typical sampling intervals were of the order of a few hundred $\mu$s. We call the last $20\%$ of each simulation the \emph{equilibrium interval}, and all further analysis is carried out with data from this interval. Here the deviation from the expected virial $\langle K \rangle_v = n/(n+2) E$ was $<1\%$ for all systems.

\section{\label{sec:calculation}Heat capacity calculation}
The heat capacity can be calculated directly from MD data in a $\bm{NVE}$ ensemble in two ways. The first is an analytically-exact formula involving means of probability distributions related to the system kinetic energy, and the second is an approximation involving the variance in system kinetic energy. Agreement of calculated values with Eq.~(\ref{eq:sp}) then gives an indication that energy is being equipartitioned in the system.

\subsection{Exact formula for the specific heat}
An exact formula for the specific heat in a $\bm{NVE}$ ensemble is obtained by taking an energy derivative of the so-called microcanonical temperature, which in 1D gives:\cite{Rugh1998}
\begin{equation}
C_V = \frac{k_B}{N} \left ( 1 - \frac{(N-4) \langle 1/K^2 \rangle }{(N-2)\langle 1/K \rangle^2 } \right )^{-1}.
\label{eq:Cmc}
\end{equation}
This formula is related to the number of degrees of freedom in the system phase space. Perturbing a chain by asymmetric edge impulses results in $N$ independent grain kinetic energies; however, it is possible to reduce the number of independent degrees of freedom by, for example, imposing periodic boundary conditions\cite{Shirts2006} or by symmetrically perturbing the system (initial velocity perturbations at both chain ends of equal magnitude, directed into the chain). Perturbing the system in this way results in a mirror-reflection symmetry to be induced, thus halving the degrees of freedom, and in turn affecting the specific heat.

In particular, symmetric perturbations result in a system with \emph{only} $N/2$ independent grain kinetic energies if $N$ is even, and \emph{only} $(N-1)/2$ if $N$ is odd (since the central grain never moves in this case). The microcanonical specific heat must be modified to account for this loss in degrees of freedom. In Appendix~\ref{sec:App_A} we present the derivation of the correct expression for the specific heat. For even-$N$ the result is
\begin{equation}
C_{V,\mathrm{even}} = \frac{2k_B}{N} \left ( 1 - \frac{(N-8)}{(N-4)} \frac{\langle 1/K^2 \rangle}{\langle 1/K \rangle^2} \right )^{-1},
\label{eq:C_even}
\end{equation}
and for odd-$N$ it is
\begin{equation}
C_{V,\mathrm{odd}} = \frac{2k_B}{(N-1)} \left ( 1 - \frac{(N-9)}{(N-5)} \frac{\langle 1/K^2 \rangle}{\langle 1/K \rangle^2} \right )^{-1}.
\label{eq:C_odd}
\end{equation}
While these equations seem little changed compared to the original Eq.~(\ref{eq:Cmc}) with full degrees of freedom, we show in Sec.~\ref{sec:results} section that \emph{only} these give the correct result. 

\subsection{Approximate formula for the specific heat}
Since the earliest computer simulations of liquids and gases were most often performed in the ($\bm{NVE}$) ensemble, Lebowitz et al. derived an approximation for calculating $C_V$ from fluctuations in total system kinetic energy $K$, $\langle \delta K^2\rangle\equiv\langle K^2 \rangle - \langle K\rangle^2$, given as:\cite{Lebowitz1967,Rugh1998} 
\begin{equation}
\frac{\langle\delta K^2 \rangle}{\langle K \rangle^2} = \frac{2}{N} \left( 1-\frac{1}{2 C_V}\right ), 
\label{eq:LPV}
\end{equation}
where $C_V$ is in units of $k_B$. $C_V$ is obtained simply by inverting this expression and plugging in the measured average system kinetic energy and its variance from MD data. Alternatively, substituting Eq.~(\ref{eq:sp}) into Eq.~(\ref{eq:LPV}) yields a prediction for the variance (fluctuations) in system kinetic energy when energy is equipartitioned in the system,
\begin{equation}
\langle \delta K^2 \rangle = \frac{2}{N} \left( \frac{2}{n+2}\right) \langle K \rangle ^2. 
\label{eq:systemK_fluc}
\end{equation}
In comparison to the hard-sphere case,\cite{Scalas2015} a factor of $(n+2)/2$, related to the finite Hertz potential exponent, appears here. Interestingly, Eq.~(\ref{eq:systemK_fluc}) implies that, in the equilibrium phase, $\langle \delta K^2 \rangle/\langle K \rangle ^2$ is absent of material dependence. This has been observed previously in MD simulations, see e.g. Fig.~5 of Ref.~\onlinecite{Przedborski2015}, where kinetic energy fluctuations were seen to ultimately approach the same value for chains of fixed length but comprised of different materials.

Similar to above, we consider what happens to Eq.~(\ref{eq:LPV}) when the system suffers from reduced degrees of freedom associated with symmetric edge perturbations. To this end, we set $K = \langle K \rangle + \delta K$  and expand $1/K$ in a Taylor series, proceeding in an identical fashion to that of Ref.~\onlinecite{Rugh1998}. The resulting approximate expressions obtained from truncating Eqs.~(\ref{eq:C_even}) and Eqs.~(\ref{eq:C_odd}) at $O(N^{-1})$, are analogous to Eq.~(\ref{eq:LPV}), except appropriately modified to account for the spatial symmetry of the system. The result is:
\begin{equation}
\frac{\langle \delta K^2 \rangle}{\langle K \rangle^2} = \frac{4}{N} \left ( 1 - \frac{1}{2C_{V,\mathrm{even}}} \right ) 
\label{eq:LPV_even}
\end{equation}
for even-$N$ systems, and
\begin{equation}
\frac{\langle \delta K^2 \rangle}{\langle K \rangle^2} = \frac{4}{N} \left ( 1 - \frac{N}{2(N-1)C_{V,\mathrm{odd}}} \right ) 
\label{eq:LPV_odd}
\end{equation}
for odd-$N$ systems. Note that in both Eqs.~\ref{eq:LPV_even} and~\ref{eq:LPV_odd}, $C_V$ is in units of $k_B$.

\section{\label{sec:results}Results and Discussion}
Here we show that the Hertzian chain indeed reaches an equilibrium phase at sufficiently long times by addressing the equipartitioning of energy among all grains within the equilibrium interval. Since ergodicity (defined as the equivalence of ensemble and time averages of physical observables) is not, in general, a prerequisite to establishing equilibrium,\cite{Coveney2016} we do not focus on this property here. Rather, since it is thought that the QEQ phase in Hertzian systems is ergodic, we make the assumption that the equilibrium phase is also ergodic, and establish this by more rigorous statistical test in a subsequent manuscript.\cite{Przedborski2016b} Agreement between theoretical predictions for the specific heat and numerical calculations also gives indication that ergodicity holds.

First we give results of simulations of systems with asymmetric perturbations, then we discuss the symmetrically-perturbed systems with reduced degrees of freedom.

\subsection{\label{sec:results_asym}
Asymmetric perturbations}
\begin{figure}[ht]
\centering
\includegraphics[width=0.95 \textwidth]{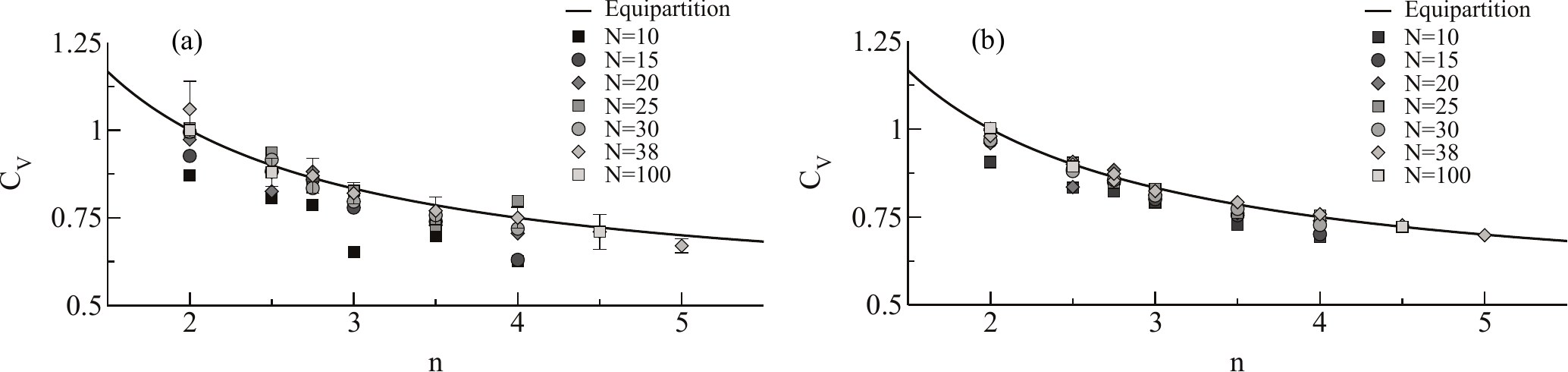}
\caption{(Color online) Specific heat capacities computed for all MD simulated systems as a function of the exponent on the potential. In (a) we present the values obtained from Eq.~(\ref{eq:Cmc}), and in (b) values obtained from inverting Eq.~(\ref{eq:LPV}). The solid line in both plots is the specific heat predicted by the generalized equipartition theorem, Eq.~(\ref{eq:sp}). The error bars are computed for $N=38$ systems in (a) as the standard deviation of calculations obtained by varying the sampling interval.}
\label{fig:fig2}
\end{figure}

To prove that energy is equipartitioned, we computed the specific heats of MD simulation data using both Eqs.~(\ref{eq:Cmc}) and (\ref{eq:LPV}). These calculated results are directly compared with $C^{\mathrm{Eq}}_V$ predicted by Eq.~(\ref{eq:sp}), shown as the solid line in both Figs.~\ref{fig:fig2}(a) and (b). It is evident that as $N$ increases, the values calculated by Eqs.~(\ref{eq:Cmc}) and (\ref{eq:LPV}) agree very well with the theory. Moreover, even for small ($N\lesssim20$) systems, the deviation from theory is no more than $\sim 10\%$ for Eq.~(\ref{eq:LPV}), and improve with additional data points in the averaging. 

The fact that the calculated specific heat agrees with $C^{\mathrm{Eq}}_V$ for $N\gg1$ provides evidence that energy is indeed equipartitioned in the Hertz chain at long enough times. This establishes that the very long-time dynamics of 1D granular chains perturbed at one end with zero dissipation is a true equilibrium phase.\cite{Avalos2011} We address whether this holds for symmetric perturbations in the next subsection. 

\subsection{\label{sec:results_sym}Symmetric perturbations}
\begin{figure}[ht]
\centering
\includegraphics[width=0.95\textwidth]{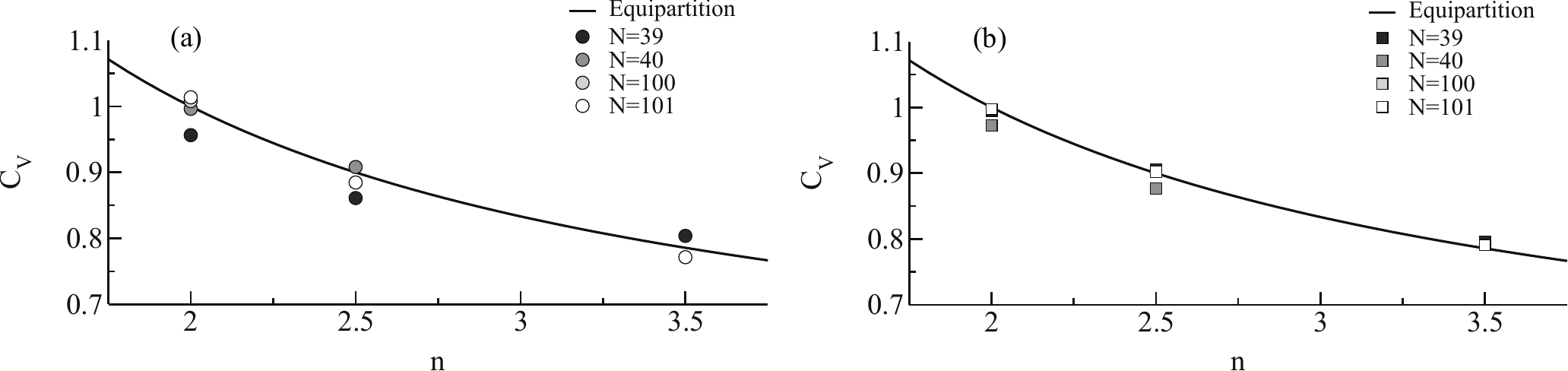}
\caption{(Color online) Specific heat capacities computed for all symmetrically-perturbed MD simulated systems as a function of the exponent on the potential. (a) Specific heat obtained from Eqs.~(\ref{eq:C_even}) and ~(\ref{eq:C_odd}); (b) Specific heat obtained from inverting Eqs.~(\ref{eq:LPV_even}) and ~(\ref{eq:LPV_odd}). The solid line is the specific heat predicted by the generalized equipartition theorem, Eq.~(\ref{eq:sp}).}
\label{fig:fig5}
\end{figure}

Here we test whether our hypothesis of equilibrium extends to systems of reduced degrees of freedom. We calculate the heat capacities of symmetrically-perturbed systems using both Eqs.~(\ref{eq:C_even}) and (\ref{eq:LPV_even}) for even-$N$ systems, and Eqs.~(\ref{eq:C_odd}) and (\ref{eq:LPV_odd}) for odd-$N$ systems. The results are shown in Fig.~\ref{fig:fig5} for two representative even-$N$ systems two representative odd-$N$ systems. 

It is clear from Figs.~\ref{fig:fig5}(a) and (b) that the calculated specific heats agree well with the values predicted by the equipartition theorem, Eq.~(\ref{eq:sp}), indicating that energy is also shared equally in symmetrically-perturbed systems. Since this is the case even for odd-$N$ systems, where the central grain never moves and therefore gets no kinetic energy, the definition of  ``equipartitioning of energy'' must be clearly defined in such systems. While energy equipartitioning is sometimes erroneously discussed in terms of energy being shared equally among all particles in a system, the equipartition theorem makes no reference to particles, but rather to the independent degrees of freedom in a system.\cite{Reif2009} Hence the fact that the specific heat, Eqs.~(\ref{eq:C_even}) and~(\ref{eq:C_odd}), agrees with the value predicted by the equipartition theorem implies that the energy is being spread out equally over all the \textit{independent degrees of freedom} at long times in Hertzian chains.

For completeness, we computed the probability distribution functions of grain velocity, grain kinetic energy, and system kinetic energy of all our MD simulations within the equilibrium interval, and note that they agree well with the approximate distribution functions recently derived\cite{Przedborski2016b,Scalas2015,Ray1991} for a 1D gas of interacting spheres in equilibrium in a microcanonical ensemble, for both asymmetrically- and symmetrically-perturbed chains. This reinforces our assertion that the chain is indeed in equilibrium at sufficiently long times. 

\section{\label{sec:conclusions}Conclusions}  
We have illustrated that the long-term dynamics of 1D Hertzian systems between fixed walls and with zero dissipation is a true equilibrium phase~\cite{Avalos2011}. We have shown that this equilibrium phase is ergodic, and that it is characterized by finite kinetic energy fluctuations that are related to the specific heat capacity in a microcanonical ensemble. For large systems, we find agreement between calculated heat capacity and values predicted by the generalized equipartition theorem, indicating that energy is equipartitioned in Hertzian chains at long times. This finally establishes that the weak interactions between SWs and SWs and boundaries in Hertzian chains drive the system dynamics at long times beyond QEQ to a thermal equilibrium phase, whose properties are predicted by equilibrium statistical mechanics. The fact that these strongly nonlinear, nonintegrable systems equilibrate very late in time, requiring extreme simulation times, agrees with assertions made in Ref.~\onlinecite{Onorato2015}, where the long-term dynamics of weakly nonlinear FPU lattices was investigated. Other work\cite{Przedborski2015} has suggested a slow (algebraic $\sim 1/t$) decay of the fluctuations to equilibrium values.

Our study has implications to the broader scientific and engineering communities. For example, quantitative analysis of the QEQ phase may now be possible with this equilibrium theory as the starting point and by employing machinery from non-equilibrium statistical physics (e.g. Boltzmann equation, linear response theory, etc.). This may allow for the early-time dynamics of the QEQ phase to be predicted and manipulated. Then being able to control the nature of the particle interactions, the system's journey to equilibrium could potentially be tuned and optimized for physical applications, such as shock disintegration.

While real granular alignments are inherently dissipative, dissipation-free versions of our systems may be possibly realized as integrated circuits and hence our results may immediately be observable in the laboratory. Moreover, we are currently making attempts to extend these ideas to include dissipation and driving. We are also working on extending these ideas to random mass and diatomic chains, as well as long-range potentials without cutoffs.

\section*{Acknowledgements}
This work was supported by a Vanier Canada Graduate Scholarship from the Natural Sciences and Engineering Research Council. 

\appendix{\label{sec:App_A} Symmetrically-imposed reduced degrees of freedom.} 
Here we extend the work of Rugh,\cite{Rugh1998} to investigate the effects that symmetrically-imposed reduced degrees of freedom have on the microcanonical heat capacity. We adopt the notation of Ref.~\onlinecite{Rugh1998}, and introduce the vector field $X_1 \equiv \mathbf{p}/(2K(p))$, where $K(p)$ is the system kinetic energy defined by $K(p) = \sum_{i=1}^N p_i^2/2$, with $p_i$ the magnitude of the grain momentum, and where we have set the grain mass $m=1$ for convenience. The vector $\mathbf{p}$ contains only the \emph{independent} translational grain momenta, i.e. $\mathbf{p} \equiv (\mathbf{p}_1,\mathbf{p}_2,\dots,\mathbf{p}_{\tilde{d}_{\bm{p}}})$ where $\tilde{d}_{\bm{p}}$ denotes the (reduced) number of degrees of freedom in momentum space.\cite{Jepps2000,Shirts2006} We next introduce the notation 
\begin{equation}
\mathrm{div}(X_1) \equiv \sum_{i=1}^{\tilde{d}_{\bm{p}}} \frac{\partial}{\partial \mathbf{p}_i} \cdot \left( \frac{\mathbf{p}_i}{2K(p)} \right),
\label{eq:div}
\end{equation}
where the dot denotes the vector dot product. The $\bm{NVE}$ temperature is related to Eq.~(\ref{eq:div}) via $T(E) = 1/\langle \mathrm{div}(X_1);E \rangle$.\cite{Rugh1998} 

When all $\mathbf{p}_i, i=1\dots N$ are independent, such as when the Hertz chain is perturbed asymmetrically, $\tilde{d}_{\bm p}=N$ and Eq.~(\ref{eq:div}) evaluates to $(N-2)/(2K(p))$. On the other hand, when the Hertz chain is perturbed symmetrically, all grain momenta are no longer independent. In particular, when $N$ is even, we have $\mathbf{p}_{N-i+1} = -\mathbf{p}_{i}$ for $i = 1\dots N/2$, hence $\tilde{d}_{\bm p}=N/2$ and
\begin{equation}
\sum_{i=1}^{\tilde{d}_{\bm p}} \frac{\partial}{\partial \mathbf{p}_i} \cdot \left( \frac{\mathbf{p}_i}{2K(p)} \right) = \sum_{i=1}^{N/2} \frac{\partial}{\partial \mathbf{p}_i} \cdot \left( \frac{\mathbf{p}_i}{2K(p)} \right) = \frac{N-4}{4K(p)} 
\label{eq:div_even},
\end{equation}
where the last equality follows from a straightforward evaluation of the preceding expression.

In the case of odd-$N$ systems that are perturbed symmetrically, we have $\mathbf{p}_{(N+1)/2} = 0$, in addition to $\mathbf{p}_{N-i+1} = -\mathbf{p}_{i}$ for $i = 1\dots (N-1)/2$, thus $\tilde{d}_{\bm p} = (N-1)/2$ and 
\begin{equation}
\sum_{i=1}^{\tilde{d}_{\bm p}} \frac{\partial}{\partial \mathbf{p}_i} \cdot \left( \frac{\mathbf{p}_i}{2K(p)} \right) = \sum_{i=1}^{\frac{(N-1)}{2}} \frac{\partial}{\partial \mathbf{p}_i} \cdot \left( \frac{\mathbf{p}_i}{2K(p)} \right) = \frac{(N-1)-4}{4K(p)}. 
\label{eq:div_odd}
\end{equation}
Expressions~(\ref{eq:div_even}) and (\ref{eq:div_odd}) then lead to, respectively, the following $\bm{NVE}$ temperatures:
\begin{eqnarray}
&&T_{\mathrm{even}}^{-1} = \left (\frac{dN-4}{4} \right) \langle 1/K(p) \rangle, \nonumber \\
&&T_{\mathrm{odd}}^{-1} = \left ( \frac{d(N-1)-4}{4} \right ) \langle 1/K(p)\rangle.
\label{eq:Tmc}
\end{eqnarray}
where the subscripts ``even/odd'' denote the parity of $N$, and we have restored the spatial dimensionality $d$.

The $\bm{NVE}$ specific heat is related to the system temperature via $1/\tilde{C}=\partial T/\partial E = -T^2 \partial T^{-1}/\partial E$. Using a relation from Rugh, Ref.~\onlinecite{Rugh1998},
we can write $1/\tilde{C} = 1 - T^2\langle \mathrm{div} \left ( \mathrm{div}(X_1)X_1\right ) ; E \rangle$, with $X_1$ defined above and with $\tilde{C}$ in units of $k_B$.  

Using the definition~(\ref{eq:div}), and after some straightforward algebra, we have
\begin{equation}
\big \langle \mathrm{div} \left ( \mathrm{div}(X_1)X_1\right ) ; E \big \rangle = \frac{(dN-4)(dN-8)}{16} \left \langle 1/K(p)^2 \right \rangle 
\end{equation}
for even-$N$ systems, and
\begin{equation}
\big \langle \mathrm{div} \left ( \mathrm{div}(X_1)X_1\right ) ; E \big \rangle =  
\frac{\big ( d(N-1)-4 \big ) \big ( d(N-1)-8 \big )}  {16} \left \langle 1/K(p)^2 \right \rangle 
\end{equation}
for odd-$N$ systems. Substituting these last expressions, as well as the $\bm{NVE}$ temperatures given in Eq.~(\ref{eq:Tmc}), into the definition of $\tilde{C}$, we then obtain the following expressions for the $\bm{NVE}$ specific heat of the symmetrically-perturbed Hertz chain:
\begin{eqnarray}
&&\tilde{C}_{\mathrm{even}}^{-1} = 1 - \frac{(N-8)}{(N-4)}\frac{\big \langle 1/K^2\big \rangle}{\big \langle 1/K \big \rangle^2}, \nonumber \\
&&\tilde{C}_{\mathrm{odd}}^{-1} = 1 - \frac{(N-9)}{(N-5)}\frac{\big \langle 1/K^2\big \rangle}{\big \langle 1/K \big \rangle^2},
\end{eqnarray}
where we have set $d=1$ in the last expressions, and $\tilde{C}_{\mathrm{even/odd}}$ are in units of $k_B$. 

When there are no symmetry restrictions (e.g. when the Hertz system is perturbed asymmetrically), the number of independent degrees of freedom in momentum-space is equivalent to the number of particles $N$. One then defines the specific heat per particle as $C_V = \tilde{C}/N$. In contrast, in  systems with reduced degrees of freedom (e.g. symmetrically-perturbed systems) it is more appropriate to define $C_V$ as the specific heat per \emph{independent degree of freedom}. Then for even $N$, $C_V = \tilde{C}_{\mathrm{even}}/(N/2)$, and for odd-$N$, $C_V = \tilde{C}_{\mathrm{odd}}/((N-1)/2)$. These are indeed the quantities predicted by the generalized equipartition theorem, Eq.~(\ref{eq:sp}).

\section*{References}


\begin{thebibliography}{0}
\bibitem{Nesterenko1983} 
V.F. Nesterenko, {\it J. Appl. Mech. Tech. Phys.} {\bf 24}(5), 733--743 (1983).

\bibitem{Nesterenko1985}
A.N. Lazaridi and V.F. Nesterenko, {\it J. Appl. Mech. Tech. Phys.} {\bf 26}(3), 405--408 (1985).

\bibitem{Nesterenko1995}
V.F. Nesterenko, A.N. Lazaridi and E.B. Sibiryakov, {\it J. Appl. Mech. Tech. Phys.} {\bf 36}(2), 166--168 (1995).

\bibitem{Sinkovits1995}
Robert S. Sinkovits and Surajit Sen, {\it Phys. Rev. Lett.} {\bf 74}(14), 2686--2689 (1995).

\bibitem{Sen1996}
Surajit Sen and Robert S. Sinkovits, {\it Phys. Rev. E} {\bf 54}(6), 6857--6865 (1996).

\bibitem{Coste1997}
C. Coste, E. Falcon and S. Fauve, {\it Phys. Rev. E} {\bf 56}(5), 6104--6117 (1997).

\bibitem{Sen1998}
Surajit Sen, Marian Manciu and James D. Wright, {\it Phys. Rev. E} {\bf 57}(2), 2386--2397 (1998).

\bibitem{Chatterjee1999}
Anindya Chatterjee, {\it Phys. Rev. E} {\bf 59}(5), 5912--5919 (1999).

\bibitem{Hinch1999}
E.J. Hinch and S. Saint{\textendash}Jean, {\it Proc. R. Soc. London, Ser. A} {\bf 455}(1989), 3201--3220 (1999).

\bibitem{Hong1999}
Jongbae Hong, Jeong-Young Ji and Heekyong Kim, {\it Phys. Rev. Lett.} {\bf 82}(15), 3058--3061 (1999).

\bibitem{Ji1999}
Jeong-Young Ji and Jongbae Hong, {\it Phys. Lett. A} {\bf 260}(1--2), 60--61 (1999).

\bibitem{Manciu1999}
Marian Manciu, Surajit Sen and Alan J Hurd, {\it Phys. A} {\bf 274}(3--4), 607--618 (1999).

\bibitem{Hascoet2000}
E. Hasco\"{e}t and H.J. Herrmann, {\it Eur. Phys. J. B} {\bf 14}(1), 183--190 (2000).

\bibitem{Sen2001}
Surajit Sen and Marian Manciu, {\it Phys. Rev. E} {\bf 64}(5), 056605 (2001).

\bibitem{Nesterenko2001}
Vitali F. Nesterenko, {\it Dynamics of hetereogeneous materials}, Springer, New York (2001).

\bibitem{Rosas2003}
Alexandre Rosas and Katja Lindenberg, {\it Phys. Rev. E} {\bf 68}(4), 041304 (2003).

\bibitem{Rosas2004}
Alexandre Rosas and Katja Lindenberg, {\it Phys. Rev. E} {\bf 69}(3), 037601 (2004).

\bibitem{Nesterenko2005} 
V.F. Nesterenko, C. Daraio, E.B. Herbold and S. Jin, {\it Phys. Rev. Lett.} {\bf 95}(15), 158702 (2005).

\bibitem{Job2007}
St{\'e}phane Job, Francisco Melo, Adam Sokolow and Surajit Sen, {\it Granul. Matter} {\bf 10}(1), 13--20 (2007).

\bibitem{Sokolow2007}
A. Sokolow, E. G. Bittle and Surajit Sen, {\it Europhys. Lett.} {\bf 77}(2), 24002 (2007).

\bibitem{Zhen2007}
Wen Zhen-Ying, Wang Shun-Jin, Zhang Xiu-Ming and Li Lei, {\it Chinese Phys. Lett.} {\bf 24}(10), 2887 (2007).

\bibitem{Herbold2009}
E.B. Herbold, J. Kim, V.F. Nesterenko, S.Y. Wang and C. Daraio, {\it Acta Mech.} {\bf 205}(1--4), 85--103 (2009).

\bibitem{Santibanez2011}
Francisco Santibanez, Romina Munoz, Aude Caussarieu, St\'ephane Job and Francisco Melo, {\it Phys. Rev. E} {\bf 84}(2), 026604 (2011).
 
\bibitem{Takato2012}
Yoichi Takato and Surajit Sen, {\it Europhys. Lett.} {\bf 100}(2), 24003 (2012).

\bibitem{Vitelli2012}
Vincenzo Vitelli and Martin van Hecke, {\it Europhys. News} {\bf 43}(6), 36--39 (2012).

\bibitem{Sen2001b} 
Surajit Sen, Felicia S. Manciu and Marian Manciu, {\it Phys. A} {\bf 299}(3--4), 551--558 (2001).

\bibitem{Nakagawa2003}
M. Nakagawa, J. H. Agui, D. T. Wu, and D. V. Extramiana, {\it Granul. Matter} {\bf 4}, 167 (2003).

\bibitem{Sokolow2005}
A. Sokolow, J.M. Pfannes, R.L. Doney, M. Nakagawa, J.H. Agui, S. and Sen, {\it App. Phys. Lett.} {\bf 87}(25), 254104 (2005).

\bibitem{Hong2005}
Jongbae Hong, {\it Phys. Rev. Lett.} {\bf 94}(10), 108001 (2005).

\bibitem{Doney2006}
Robert Doney and Surajit Sen, {\it Phys. Rev. Lett.} {\bf 97}(15), 155502 (2006).

\bibitem{Melo2006}
F. Melo, S. Job, F. Santibanez, and F. Tapia, {\it Phys. Rev. E} {\bf 73}, 041305 (2006); 
Lautaro Vergara, {\it Phys. Rev. E} {\bf 73}(6), 066623 (2006). 

\bibitem{Doney2009}
Robert L. Doney, Juan H. Agui, and Surajit Sen, {\it J. App. Phys.} {\bf 106}(6), 064905 (2009).

\bibitem{Daraio2006} 
C. Daraio, V. Nesterenko, E. Herbold and S. Jin, {\it Phys. Rev. E.} {\bf 73}(2), 026610 (2006).

\bibitem{Job2009}
St\'ephane Job, Francisco Santibanez, Franco Tapia, Francisco and Melo, {\it Phys. Rev. E} {\bf 80}, 025602 (2009).

\bibitem{Theocharis2009}
G. Theocharis, M. Kavousanakis, P.G. Kevrekidis, Chiara Daraio, Mason A. Porter, and I.G. Kevrekidis, {\it Phys. Rev. E} {\bf 80}, 066601 (2009).

\bibitem{Boechler2010}
N. Boechler, G. Theocharis, S. Job, P. Kevrekidis, M. Porter, and C. Daraio, {\it Phys. Rev. Lett.} {\bf 104}(24), 244302 (2010).

\bibitem{Theocharis2010}
G. Theocharis, N. Boechler, P.G. Kevrekidis, S. Job, Mason A. Porter, and C. Daraio, {\it Phys. Rev. E} {\bf 82}, 055604 (2010).

\bibitem{Breindel2011}
Alexander Breindel, Diankang Sun, and Surajit Sen, {\it Appl. Phys. Lett.} {\bf 99}(6), 063510 (2011).

\bibitem{Przedborski2015b}
Michelle A. Przedborski and Thad A. Harroun and Surajit Sen, {\it Appl. Phys. Lett.} {\bf 107}, 244105 (2015).

\bibitem{QEQ} 
Surajit Sen, T.R. Krishna Mohan and Jan M.M. Pfannes, {\it Phys. A} {\bf 342}(1--2), 336--343 (2004);
T.R. Krishna Mohan and Surajit Sen, {\it Pramana} {\bf 64}(3), 423--431 (2005).

\bibitem{Sen2005} 
Surajit Sen, Jan M.M. Pfannes and T.R. Krishna Mohan, {\it J. Korean Phys. Soc.} {\bf 46}, 577--579 (2005).

\bibitem{Job2005} 
St\'ephane Job, Francisco Melo, Adam Sokolow and Surajit Sen, {\it Phys. Rev. Lett.} {\bf 94}(17), 178002 (2005).

\bibitem{Avalos2007} 
Edgar \'{A}valos, Robert L. Doney and Surajit Sen, {\it Chinese J. Phys.} {\bf 45}(6-II), 666--674 (2007).

\bibitem{Avalos2011} 
Edgar \'{A}valos, Diankang Sun, Robert L. Doney and Surajit Sen, {\it Phys. Rev. E} {\bf 84}(4), 046610 (2011).

\bibitem{SSWs} 
Marian Manciu, Surajit Sen and Alan J. Hurd, {\it Phys. Rev. E} {\bf 53}(1), 016614 (2000);
Felicia S. Manciu and Surajit Sen, {\it Phys. Rev. E} {\bf 66}(1), 016616 (2002).

\bibitem{Avalos2014} 
Edgar \'{A}valos and Surajit Sen, {\it Phys. Rev. E} {\bf 89}(5), 053202 (2014).

\bibitem{Sen2008} 
S. Sen, J. Hong, J. Bang, E. Avalos and R. Doney, {\it Phys. Rep.} {\bf 462}(2), 21--66 (2008).

\bibitem{Przedborski2015} 
Michelle Przedborski, Thad A. Harroun and Surajit Sen, {\it Phys. Rev. E} {\bf 91}(4), 042207 (2015).

\bibitem{Han2014} 
Ding Han, Matthew Westley and Surajit Sen, {\it Phys. Rev. E} {\bf 90}(3), 032904 (2014).

\bibitem{Onorato2015} 
Miguel Onorato, Lara Vozella, Davide Proment, and Yuri V. Lvov, {\it Proc. Natl. Acad. Sci. U.S.A.} {\bf 112}(14), 4208--4213 (2015).

\bibitem{Przedborski2016} 
Michelle Przedborski, Surajit Sen, and Thad A. Harroun, Fluctuations in Hertz chains at equilibrium. Unpublished. arXiv:1605.08970 (2016).

\bibitem{Fermi1955} 
Enrico Fermi, J. Pasta, and S. Ulam, {\it Los Alamos Report LA-1940} {\bf 978} (1955).

\bibitem{Lebowitz1967} 
J.L. Lebowitz, J.K. Percus and L. Verlet,
{\it Phys. Rev.} {\bf 153}(1), 250 (1967).

\bibitem{Tolman1918} 
Richard Tolman, {\it Phys. Rev.} {\bf 1}(4), 261--275 (1918).

\bibitem{Hertz1882} 
Heinrich Hertz, {\it J. Reine Angew. Math.} {\bf 92}, 156--171 (1882).

\bibitem{Sun2011} 
Diankang Sun, Chiara Daraio and Surajit Sen, {\it Phys. Rev. E} {\bf 83}(6), 066605 (2011).

\bibitem{Rugh1998} 
Hans Henrik Rugh, {\it J. Phys. A Math. Gen.} {\bf 31}(38), 7761 (1998).

\bibitem{Coveney2016} 
Peter V. Coveney and Shunzhou Wan, ``On the calculation of equilibrium thermodynamic properties from molecular dynamics'', {\it Phys. Chem. Chem. Phys.} (2016).

\bibitem{Bland2013} 
Martin Bland, {\it PLoS ONE} {\bf 8}(10), 1--5 (2013).

\bibitem{Shirts2006} 
Randall B. Shirts, Scott R. Burt and Aaron M. Johnson, {\it J. Chem. Phys.} {\bf 125}(16), 164102 (2006). 

\bibitem{Reif2009} 
Frederick Reif,  {\it Fundamentals of statistical and thermal physics}. Illinois; Waveland Press (2009).

\bibitem{Przedborski2016b} 
Michelle Przedborski, Surajit Sen, and Thad A. Harroun, Distribution functions in 1D gases of interacting spheres in a microcanonical ensemble. In preparation. (2016).

\bibitem{Scalas2015} 
Enrico Scalas, Adrian T. Gabriel, Edgar Martin and Guido Germano, {\it Phys. Rev. E} {\bf 92}(2), 022140 (2015).

\bibitem{Ray1991} 
John R. Ray and H. W. Graben, {\it Phys. Rev. A} {\bf 44}(10), 6905--6908 (1991). 

\bibitem{Jepps2000} 
O. G. Jepps, G. Ayton, and D. J. Evans, {\it Phys. Rev. E}, {\bf 62}(4), pp. 4757 (2000).

\end{thebibliography}
\end{document}